%% file: main.tex
\documentclass[letter]{aa} 
\pdfoutput=1 
\usepackage{hyperref}
\usepackage{graphicx}
\usepackage{txfonts}
\usepackage{amsmath}
\usepackage{float}
\usepackage{physics}
\usepackage{siunitx}
\usepackage{amssymb}
\usepackage{bm}
\usepackage{dcolumn}
\usepackage{siunitx}

\begin{document} 


\input{0Title_authors_abstract_keywords}
\input{1Introduction.tex}
\input{2Data-MONDorbits.tex}
\input{3PoR.tex}
\input{5Discussion_and_Conclusion.tex}


\begin{acknowledgements}
We thank Bob Sanders (and indirectly M. Milgrom) for sharing his knowledge on MOND and very  useful discussions. We are also grateful to the referee for their very constructive report. We~acknowledge~financial~support from a Spinoza prize.
We have made use of data from the European Space Agency (ESA) mission
{\it Gaia} (\url{https://www.cosmos.esa.int/gaia}), processed by the {\it Gaia}
Data Processing and Analysis Consortium (DPAC,
\url{https://www.cosmos.esa.int/web/gaia/dpac/consortium}). Funding for the DPAC
has been provided by national institutions, in particular the institutions
participating in the {\it Gaia} Multilateral Agreement. 

Throughout this work, we have made use of the following packages: \texttt{astropy} \citep{Astropy},
          \texttt{vaex} \citep{vaex2018},
          \texttt{SciPy} \citep{2020SciPy-NMeth},
          \texttt{matplotlib} \citep{matplotlib},
          \texttt{NumPy} \citep{Numpy},
          \texttt{AGAMA} \citep{Vasiliev2019} and Jupyter Notebooks \citep{JupyterNotebook}.
\end{acknowledgements}

\bibliography{bibliography}
\bibliographystyle{aa} 


\end{document}

%% file: 0Title_authors_abstract_keywords.tex

\title{Testing MOND using the dynamics of nearby stellar streams}


 \author{{Orlin Koop\thanks{e-mail: koop@astro.rug.nl}}
          \and 
          {Amina Helmi}
          }

   \institute{Kapteyn Astronomical Institute, University of Groningen, Landleven 12, NL-9747 AD Groningen, the Netherlands}

   \date{Received xxxx; accepted yyyy}

\abstract
   {The stellar halo of the Milky Way is built up, at least in part, from debris from past mergers. Stars from such merger events define substructures in phase-space, for example in the form of streams, which are groups of stars moving on similar trajectories. The nearby Helmi streams discovered more than two decades ago are a well-known example. Using 6D phase-space information from the \textit{Gaia} space mission, \citet{Dodd2022a}  {have recently reported} that the Helmi streams are split into two clumps in angular momentum space. Such substructure can be explained and sustained in time if the dark matter halo of the Milky Way takes a prolate shape in the region probed by the orbits of the stars in the streams. }
   {Here, we explore the behaviour of the two clumps identified in the Helmi streams in a Modified Newtonian Dynamics (MOND) framework to test this alternative model of gravity. }
   {We perform orbit integrations of Helmi streams member stars in a  {simplified} MOND model of the Milky Way  {and using the more sophisticated \textit{Phantom of RAMSES} simulation framework.}} 
   {We find  {with both approaches that} the two Helmi streams' clumps do not retain their identity and dissolve after merely  {100} Myr. This extremely short timescale would render the detection of two separate clumps as very unlikely in MONDian gravity.}
  {The observational constraints provided by the streams, which MOND fails to reproduce in its current formulation, could potentially also be used to test other alternative gravity models.}

\keywords{Galaxy: structure — Gravitation — Galaxy: kinematics and dynamics}

\maketitle

%% file: 1Introduction.tex
\section{\label{sec:intro}Introduction}
A plentitude of empirical evidence points to mass discrepancies in the Universe. \citet{Oort1932} found the luminous matter in the Solar vicinity to be insufficient to explain the vertical motions of stars near the Sun, and \citet{Zwicky1933} reported that the velocity dispersion in (clusters of) galaxies was too high for them to stay bound by the visible mass. Later \citet{Ostriker1973}  showed that dynamically cold disks in galaxies are prone to instability unless embedded in some potential well like a halo of dark matter. Then \citet{Rubin1982} and \citet{Bosma} showed that the rotation curves of spiral galaxies remain approximately flat with increasing radius. With the baryonic mass inferred from the Big Bang Nucleosynthesis model and the observation of the cosmic microwave background and present-day inhomogeneity, the need for a boost of the growth rate with invisible mass also arose from cosmology.

\begin{figure*}
    \centering
    \includegraphics[width=\linewidth]{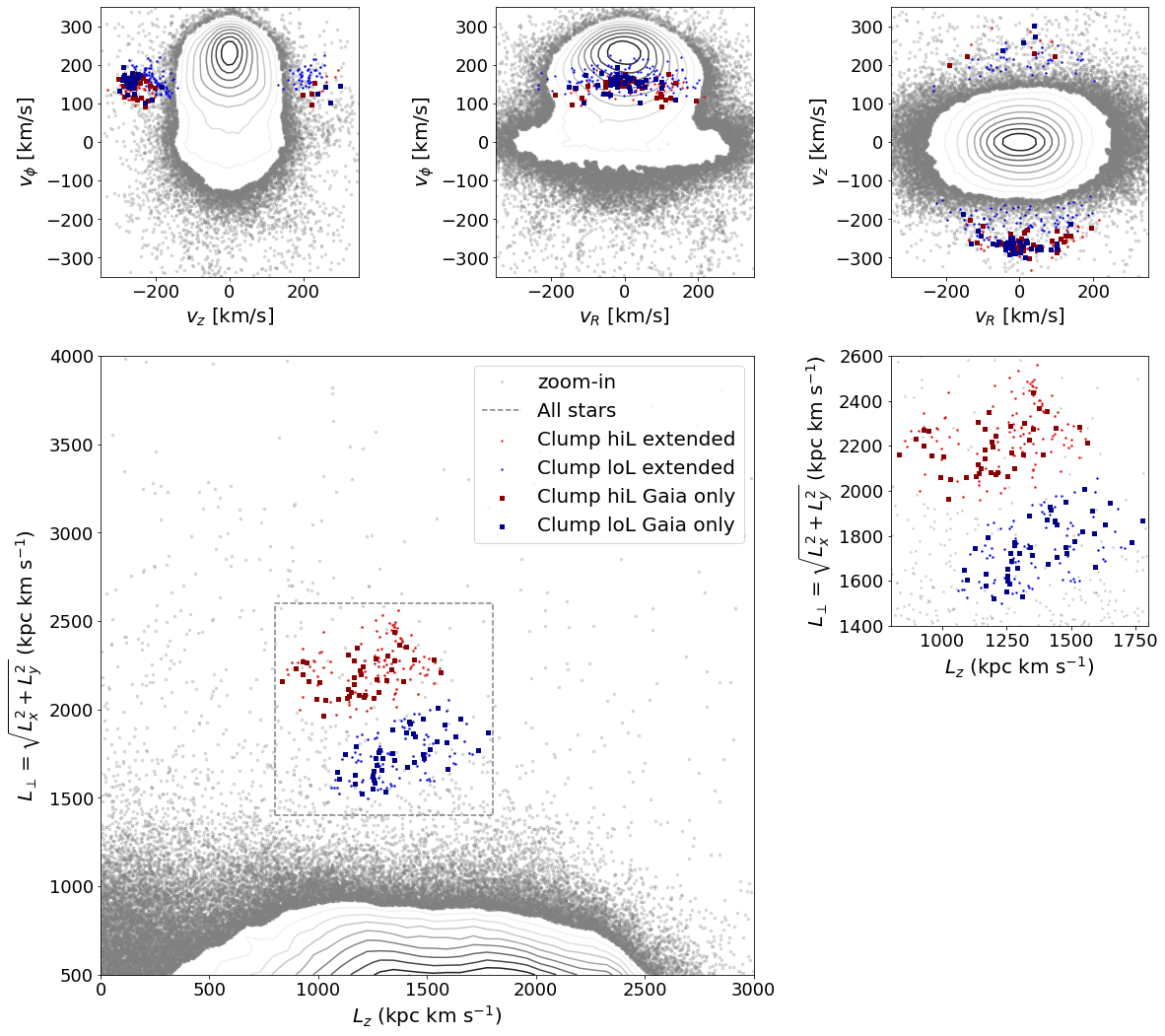}
    \caption{Helmi streams' stars within $2.5$~\si{kpc} as selected by D22 shown in colour in velocity space (top panels) and $L_z-L_{\perp}$ space (bottom-left panel and zoom-in on the right). Grey dots and contours show the full \textit{Gaia} sample within $2.5~\si{kpc}$. The darker colored squares are all Helmi streams stars with \textit{Gaia} radial velocities while the lighter colored dots show stars with ground-based radial velocities. While the Helmi streams stars are visibly clumped in $v_z-v_y$ space, the streams are clearly distributed in two distinct clumps in $L_z-L_{\perp}$ space. One with high $L_{\perp}$ (hiL, red) and one with low $L_{\perp}$ (loL, blue).}
    \label{fig:data}
\end{figure*}

The currently preferred cosmological model includes a cosmological constant ($\Lambda$) and involves significant amounts of cold dark-matter (CDM). The $\Lambda$CDM model is supported by a wealth of observational data over many different scales. It does however face some challenges  {on small scales} such as the predicted excess of dark satellites (or subhalos), i.e. the `missing satellite problem' \citep[see e.g.][]{Boylan-Kolchin2011,Bullock2017,Pawlowski2017}, or the
`plane-of-satellites problem' where the observed spatial distribution of the satellites of Milky Way seems to be too planar in comparison to predictions from simulations (see e.g \citet{Libeskind2005}, \citet{Sawala2023}, \citet{Garavito-Camargo2023} and \citet{Sales} for a review and possible solutions and outstanding problems on the dwarf galaxy scale). Another challenge is the `radial acceleration relation' (RAR), where the rotation curve of many galaxies can be predicted from the Newtonian gravitational acceleration alone in a universal way \citep[see e.g.][and references therein]{Banik2022}.

An alternative way of looking at the hidden mass hypothesis is that the problem signals a breakdown of Newtonian dynamics in the weak acceleration regime, defined by a new constant $a_0$, of the order of $10^{-10}$~\si{m.s^{-2}}, found to be appropriate to the tiny accelerations encountered in galaxies beyond the Solar radius \citep{Milgrom1983,GaiaCollaboration2021b}. MOND (MOdified Newtonian Dynamics) is able to  {predict} the observation of flat rotation curves, the RAR and the baryonic Tully-Fisher relation --a scaling relation between the luminosity and the circular velocity of disk galaxies--   {\citep{Famaey2012,Milgrom1983b,Lelli2016,McGaugh2016}}. However, MOND runs into challenges on more cosmological scales, even needing additional
dark matter to explain certain observations of galaxy clusters  {\citep{Sanders1999,Skordis2021}}.

With the increasing amount of data on our own Galaxy, especially from the \textit{Gaia} mission \citep{GaiaCollaboration2016},
more tests of MOND on the Galactic scale can be performed. However, due to the nonlinearity of MOND (i.e.~forces are not additive), attempts have been somewhat limited thus far, except for instance the work done by \cite{Read2005}. Some other examples include using the kinematics of stars to model the rotation curve and mass distribution in the Galaxy 
\citep{Banik2018,Lisanti2019,Dai2022,Finan-Jenkin2023,Zhu2023,SylosLabini2023} and the modeling of stellar streams  {\citep{Thomas2017,Thomas2018,Kroupa2022}}  {using the \textit{Phantom of RAMSES} patch \citep[PoR, ][]{Lughausen2015}.}
Stellar streams are the products of the tidal stripping of satellites like globular clusters or dwarf galaxies. 
These streams are a great probe of the galactic mass distribution and potential. Not only are their characteristics 
dependent on the overall gravitational potential, but the presence of dark satellites (often referred to as subhalos, predicted in the thousands in the case of cold dark matter) can result in gaps, spurs or other substructures in streams  {\citep{Ibata2002,Johnston2002,Erkal2015}.} Such features would have to be attributed to interactions with baryonic structures in the case of MOND.  Furthermore, tidal streams resulting from globular clusters in MOND will experience an altered internal potential, leading to an asymmetry between the leading and the trailing tidal tails \citep{Thomas2018}, which is generally not expected in Newtonian gravity  \citep[see][for more discussion]{Dehnen2004,Pearson2017}.

Among the stellar streams identified thus far are the Helmi streams crossing the Solar neighbourhood \citep{Helmi1999a,Koppelman2019}. They are thought to be the debris of a massive dwarf galaxy with a stellar mass of $M_{*}\sim10^8~M_{\odot}$, that was accreted approximately 5--8~\si{Gyr} ago. 
Its stars have phase-mixed, implying that they do not define spatially tight structures but that the streams' stars cross the Solar vicinity on different phases of their orbit. They define clumps in velocity space as well as integrals of motion space (see Fig.~\ref{fig:data}). The latter is defined by (quasi-)conserved quantities along the orbit of the stars, 
such as their energy and angular momenta. These can be used to track dynamically groups of stars of similar origin \citep{HdZ}. \cite{Dodd2022a} (hereafter D22) have shown that the Helmi streams in fact define two distinct
clumps in angular-momentum space ($L_z-L_{\perp}$, where $L_{\perp} = \sqrt{L_x^2+L_y^2}$). The existence of these two subclumps has been confirmed by a clustering algorithm using both \textit{Gaia} EDR3 \citep{Lovdal2022,Ruiz-Lara2022} and the more recent \textit{Gaia} DR3 dataset \citep{Dodd2023}. The stellar populations of these two clumps are indistinguishable, which means that they must have originated in the same system. D22 demonstrate that the two clumps in ($L_z-L_{\perp}$) space are able to survive in time as such if the Galactic dark matter halo has a prolate shape with axis ratio $q_{\rho} \sim 1.2$. In this case, one of the clumps is placed on the
 $\Omega_{\phi}$:$\Omega_z$=1:1 orbital resonance.
 
The need for a flattened shape of the dark matter halo that does not follow the distribution of baryons, prompted us to want to explore the dynamics of the Helmi streams in the context of MOND. In particular, we study here the behaviour of the two clumps in $L_z-L_{\perp}$ space in a MOND potential that is constrained to follow the known distribution of baryons in the Milky Way.

In Section \ref{sec:data} we summarize the selection criteria on Helmi streams stars from D22. In Section \ref{sec:methods} we explain the MOND framework and how we apply it to study the Helmi streams  {with simple orbit integrations}.  {Section \ref{sec:PoR} shows our results using the Phantom of RAMSES N-body code. We present a discussion and our conclusions} in Section \ref{sec:Discussion and Conclusion}.

%% file: 2Data-MONDorbits.tex
\section{Data}
\label{sec:data}

We follow D22 in selecting Helmi streams stars from data provided by \textit{Gaia} EDR3 \citep{GaiaCollaboration2021}. This mission has provided 6D information for 7,209,831 of their $\sim1.7$ billion sources. After applying two quality cuts, namely \texttt{parallax\_over\_error}$>5$ and RUWE~$<1.4$, there remain $4,496,187$ sources within $2.5$~\si{kpc} of the Sun. This sample was extended by including radial velocities
from crossmatches with spectroscopic surveys, namely APOGEE DR16 \citep{Ahumada2020}, LAMOST DR6 \citep{Cui2012}, Galah DR3 \citep{Buder2021} and RAVE DR6 \citep{Steinmetz2020}. This results in a sample of $7,531,934$ sources. These stars have been corrected for the parallax zero point offset \citep[$-17~\si{\mu as}$,][]{Lindegren2021}, solar motion \citep[with $(U,V,W)_{\odot}=(11.1,12.24,7.25)~\si{km.s^{-1}}$,][]{Schonrich2010} and motion of the local standard of rest \citep[$v_{\mathrm{LSR}}=232.8~\si{km.s^{-1}}$,][]{McMillan2017}, to transform their coordinates and proper motions to Galactocentric Cartesian coordinates  assuming $R_{\odot}=8.2~\si{kpc}$ and $z_{\odot}=0.014~\si{kpc}$ \citep{McMillan2017,Binney1997}.
We calculate angular momenta, $L_z$ and $L_{\perp}$ with the sign of $L_z$ flipped to that $L_z>0$ for prograde orbits. 

To select stars belonging to the Helmi streams, D22 calculate the orbital energy of the stars with \texttt{AGAMA} \citep{Vasiliev2019} 
in an axi-symmetric potential consisting of a stellar thin and thick disk, an HI gas disk, a molecular gas disc, a spherical bulge and an NFW halo, which is spherical by default \citep{McMillan2017}. D22 define ellipses to select two clumps in $L_z-L_{\perp}$ after removing less-bound stars  (with  $E<-1.2\times 10^5$~\si{km^2.s^{-2}}) which are unlikely to be members of the streams. The clump at higher $L_{\perp}$ contains 154 stars and the clump at lower $L_{\perp}$ contains 130 stars. For more information, see Section 2 in D22. 
Figure \ref{fig:data} shows the kinematics of the Helmi streams stars as well as their 
distribution in $L_z-L_{\perp}$ space, where the gap between the clumps is very apparent for the stars with \textit{Gaia} radial velocities because of their much smaller measurement uncertainties.

\section{ {MOND}}
\label{sec:methods}

To investigate the behaviour of the Helmi streams in the context of MOND, we  {use} the AQUAL formulation of MOND \citep{Bekenstein1984}. The original formulation of MOND states that the MOND gravitational acceleration $\bm{g}$ is given by \citep{Milgrom1983}:
\begin{equation}
\mu\left(\dfrac{g}{a_0}\right)\bm{g}=\bm{g}_{\rm N},\label{eq:origMOND}
\end{equation}
where $\bm{g}_{\rm N}$ is the Newtonian gravitational acceleration and $\mu$ is the interpolation function between the `deep-MOND' regime at low accelerations and the Newtonian regime at high accelerations. At these accelerations, one must recover the observed behaviour, namely that the total gravitational attraction approaches $g_{\rm N}$, and therefore $\mu(x)\rightarrow1$ for $x\gg1$. On the other hand, for low accelerations  $g\rightarrow \sqrt{g_{\rm N} a_0}$, where $a_0=1.2\times10^{-10}$~\si{m.s^{-1}} is Milgrom's acceleration constant, and so $\mu(x)\rightarrow x$ for $x\ll1$.
In this paper, we employ the following widely used interpolating function: 
\begin{equation}
\mu(x)=\dfrac{x}{1+x},
\end{equation}
which has been shown to provide good fits in the intermediate to weak gravity regimes of galaxies \citep{Famaey2005}.

In the AQUAL formulation of MOND, Eq.~(\ref{eq:origMOND}) results from a modification of the gravitational action (or Lagrangian) at a classical level, which also changes the Poisson equation. The AQUAL Poisson equation is:
\begin{equation}
    \bm{\nabla}\left[\mu\left(\dfrac{|\bm{\nabla}\Phi|}{a_0}\right)\bm{\nabla}\Phi\right]=4\pi G\rho_{\rm b},\label{eq:QUMONDPoisson}
\end{equation}
where $\Phi$ is the MOND gravitational potential, and $\rho_{\rm b}$ is the baryonic matter density. A general solution for the gravitational force arising from the MOND potential can be written as:
\begin{equation}
  \mu\left(\dfrac{g}{a_0}\right)\bm{g}=\bm{g}_{\rm N} + \bm{S},\label{eq:solwithS}
\end{equation}
where $\bm{S}$ is a curl field so that $\bm{\nabla}\cdot \bm{S}=0$, and $\bm{g}_{\rm N}=-\nabla\Phi_{\rm N}$, where $\Phi_{\rm N}$ is the Newtonian gravitational potential. Hence, the original formulation of MOND  (i.e. Eq.~\ref{eq:origMOND}) would only be applicable in systems where $\bm{S}$ vanishes. This has been shown to hold in highly symmetric and/or one-dimensional systems, i.e. when one can write $|\bm{\nabla}\Phi_{\rm N}|=f(\Phi_{\rm N})$ for some smooth function $f$. An important example for which this holds are flattened systems where the isopotential surfaces are locally spherically symmetric, such as Kuzmin disks and disk-plus-bulge generalizations thereof \citep{Brada1995}. Indeed, for a Kuzmin disk of mass $M$, it can be shown that $f(\Phi_{\rm N})=\Phi_{\rm N}^2/MG$.

\begin{figure*}
    \centering
    \includegraphics[width=\linewidth]{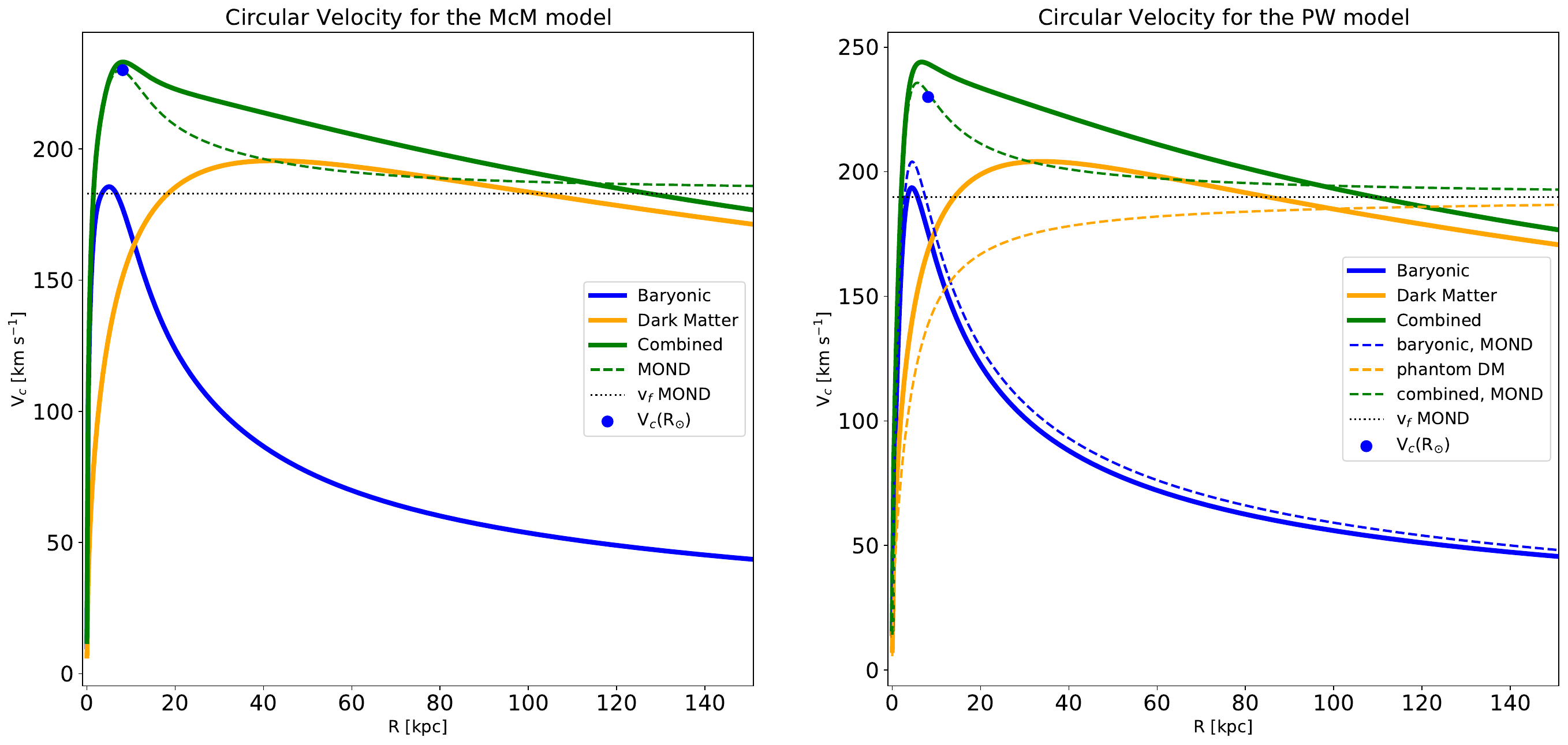}
    \caption{The rotation curves for the  {McM model} (left panel) and the  {PW model} (right panel). Solid lines show the components for the Newtonian framework. All dashed lines correspond to the MOND framework. The black horizontal dotted line shows the MONDian asymptotic velocity of the velocity curve, $v_{\rm f}=\sqrt[4]{GM_{\mathrm{bary}}a_0}$, where $M_{\mathrm{bary}}$ is the total baryonic mass. The dot shows the circular velocity at the solar radius from \cite{McMillan2017}. The baryonic models in the Newtonian and MOND framework have the same parameters, except for the increased disk mass in the MOND framework, needed to fit $V_{\rm c}(R_{\odot})$.}
    \label{fig:vcPW}
\end{figure*}

\subsection{Milky Way Models}

To derive a Milky Way potential in the MOND framework, and hence the force field necessary to integrate the orbits of the Helmi streams' stars, we focus on the baryonic components of our Galaxy, namely its disk and bulge. For simplicity, we use the descriptions of these components from two widely employed models that we describe below. In this way we extend the pioneering work of \cite{Read2005} by using more complex potentials than the Kuzmin disk.  {We note that \citet{Brada1995} have shown for disk rotation curves that the original approximation given by Eq.~(\ref{eq:origMOND}) holds beyond $R=5$~kpc in Kuzmin and exponential disks, and that this equation provides a good approximation to the rotation curves for bulge plus disk potentials \citep{Milgrom1986}. This means we can use the constraint of the circular velocity at $R_{\odot}=8.2~\si{kpc}$ to fit our MOND models.}

 {We focus first on the McMillan model} (McM model), which stems from the baryonic component from \cite{McMillan2017}. In order to match the measured circular velocity at the position of the Sun, we needed to increase the surface mass density of the thin disk component slightly by $10\%$ from $\Sigma_{\rm d}= 9\times10^8M_{\odot}$ to $1\times10^{9}M_{\odot}$. The rotation curve provided by this model can be seen in the left panel of Fig. \ref{fig:vcPW} and compared to the original Newtonian McM model.  {Note that we have modified the dark matter halo of the latter model to have $q_{\rho}=1.2$, following D22. This however, does not have any effect on the rotation curve in the disk plane.}

 {We also explore the Price-Whelan model }(PW model), which consists of a disk and a bulge, adapted from the baryonic component of the \texttt{MilkyWayPotential} in \texttt{gala v1.4} \citep{Price-Whelan2017,Price-Whelan2021}, building on the disk model from \cite{Bovy2015}. The disk has a Miyamoto-Nagai potential \citep{Miyamoto1975}:
\begin{equation}
    \Phi_{\rm MN}(R,z) = \dfrac{-GM_{\rm d}}{\sqrt{R^2+\left(a_{\rm MN}+\sqrt{z^2+b_{\rm MN}^2}\right)^2}},\label{eq:MNPot}
\end{equation}
with $R=\sqrt{x^2+y^2}$, $M_{\rm d}$~=~$7.5\times 10^{10}M_{\odot}$, $a_{\rm MN}$~=~3~kpc and $b_{\rm MN}$~=~280~pc. We again have increased the mass of the disk in this model from its default value by $\sim$~15\%.  The bulge has a potential following a \cite{Hernquist1990} profile:
\begin{equation}
    \Phi_{\rm H}(r) = \dfrac{-GM_{\rm b}}{r+a_{\rm H}},\label{eq:HernquistPot}
\end{equation}
with $r=\sqrt{R^2+z^2}$, $M_{\rm b}$ =  $5\times10^{9}M_{\odot}$ and $a_{\rm H}$  = 1~kpc.

In Newtonian gravity, the default Price-Whelan model has a dark matter halo that follows a flattened NFW potential \citep{Navarro1997}:
\begin{equation}
    \Phi_{\mathrm{DM, PW}}=\dfrac{-GM_{\rm h}}{\tilde{r}}\ln\left(1+\dfrac{\tilde{r}}{r_{\rm a}}\right),\label{eq:PhiDMPW}
\end{equation}
where $\tilde{r}=\sqrt{R^2+(z/q)^2}$, $q=0.95$ is the flattening in the $z$-direction, $M_{\rm h}=7\cdot10^{11}M_{\odot}$ is the halo mass and $r_{\rm a}=15.62~\si{kpc}$ is the scale radius.

The left panel of Fig. \ref{fig:vcPW} shows the rotation curves of the  {McM} model in both the Newtonian (i.e.~including the dark matter halo, solid) and MOND (dashed) frameworks. We calculated circular velocities as $v_{\rm c}(r)=\sqrt{rF(r)}$ in the disk plane $z=0$, where we use Eq.~(\ref{eq:gmond}) for the MOND forces.

\begin{figure*}[ht!]
    \centering
    \includegraphics[width=\linewidth]{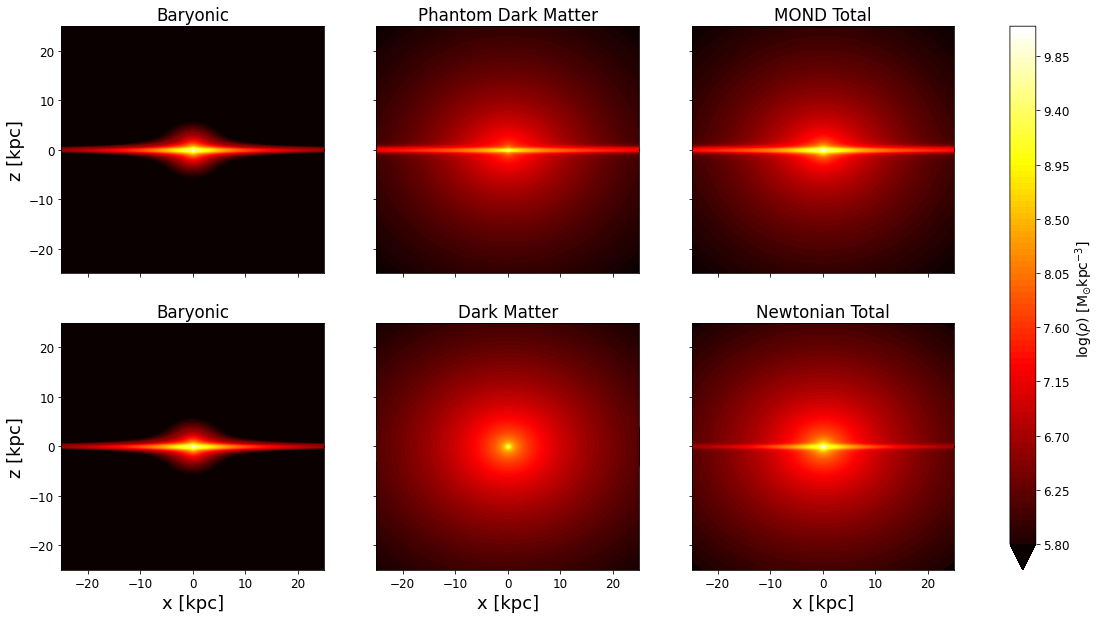}
    \caption{Density profiles of the baryonic [left column], (phantom) dark matter [middle column] and total density for the PW model, showing how the phantom dark matter is constrained to follow the shape of the disk baryonic distribution.}
    \label{fig:densities}
\end{figure*}

 {The right panel shows the results for the PW model, where we also} show the contribution from the phantom dark matter density. 
 {This} \textit{phantom dark matter density} is the effective density of dark matter that would have caused the MOND force field in Newtonian gravity, i.e.
\begin{equation}
    \bm{\nabla}^2\Phi=4\pi G(\rho_{\mathrm{b}} + \rho_{\mathrm{ph}}),
\end{equation}
where $\rho_\mathrm{b}$ is the baryonic density, and $\rho_{\mathrm{ph}}$ is the phantom dark matter density. Therefore, if the Newtonian potential is known we can directly infer the functional form of the phantom dark matter density:
\begin{equation}
    \rho_{\mathrm{ph}}=\dfrac{\bm{\nabla}\cdot((\nu(|\bm{\nabla}\Phi_{\rm N}|/a_0)-1)\bm{\nabla}\Phi_{\rm N})}{4\pi G}.\label{eq:phantom}
\end{equation}
Here 
\begin{equation}
	\nu(x) = \dfrac{\sqrt{1+4/x}+1}{2}
\end{equation}
is a function such that one can write 
\begin{equation}
\label{eq:gmond}
\bm{g}=\nu(g_{\rm N}/a_0)\bm{g}_{\rm N},
\end{equation}
which in terms of Eq.~(\ref{eq:origMOND}) means that $\nu(y)=1/\mu(x)$, where $y=x\mu(x)$ for ${\bm S} = {\bm 0}$ \citep{Famaey2012}.

Figure \ref{fig:densities} compares the density of the PW model in the $x-z$-plane for the Newtonian (bottom row) and MOND (top row) frameworks. The contours reveal how the phantom dark matter follows the shape of the disk at low heights above the plane, but is spherically symmetric farther away.

\subsection{The EFE}
An important characteristic of MOND, due to its nonlocality, is that the external field acting on a system can have a significant effect \citep[see e.g.][]{Milgrom1983,Milgrom2010,Banik2018}. In our case, when studying the individual orbits of Helmi streams stars in the Milky Way, this external field effect (EFE) could arise from the presence of dwarf galaxies like Sagittarius (Sgr) or the Large Magellanic Cloud (LMC), or due to Andromeda (M31), the largest galaxy closest to the Milky Way. To study the importance of accounting for the EFE due to these objects, we calculate where an external field produces a force equal to the internal field of the Milky Way. This happens at a radius of:
\begin{equation}
    R_{\mathrm{EFE}} = \dfrac{\sqrt{GM_{\mathrm{bary, MW}}a_0}}{g_{\mathrm{ext}}},
\end{equation}
where $M_{\mathrm{bary, MW}}$ is the total baryonic mass of the Milky Way in the MOND framework, and $g_{\mathrm{ext}}$ is the MOND gravitational acceleration due to the external object.
For an object with total baryonic mass $M_{\mathrm{bary}}$ and radius $r$, the rotational velocity flattens at a plateau with a final velocity $v_{\rm f}=~\sqrt[4]{GM_{\mathrm{bary}}a_0}$, so the MOND gravitational acceleration felt at a distance $d\gg r$ from such an object is $g_{\mathrm{ext}}=v_{\rm f}^2/d$.

For M31, $v_{\rm f}\approx225$~\si{km.s^{-1}} \citep{Carignan2006} and $d\approx770$~\si{kpc}, so $g_{\mathrm{ext}}=0.02a_0$, which is well within the deep MOND limit. Therefore, $R_{\mathrm{EFE}} \sim  450 - 490$~\si{kpc} for the external field effect from Andromeda on the Milky Way for the McM and PW models.
In our orbit integrations Helmi streams stars reach an apocenter  of 21 kpc, which is much smaller than $R_{\mathrm{EFE}}$, and hence the orbits should not be affected significantly by the external field effect due to M31 \citep{Banik2020}. Additionally,  it has been shown \cite{Zhao2013}  that M31 has remained farther away than $600~\si{kpc}$ from the Milky Way in the past $\sim8~\si{Gyr}$, meaning that the Helmi streams have been orbiting the Galaxy without considerable influence from M31 since their accretion, which has been estimated to have taken place 5 - 8 Gyr ago \citep{Koppelman2019}.  {This conclusion is supported by \cite{Oria2021} who find that the external field effect of M31 and also of the Virgo cluster can be neglected (for objects orbiting) in the inner regions of the Milky Way.}

\begin{figure*}[ht!]
    \centering
    \includegraphics[width=\linewidth]{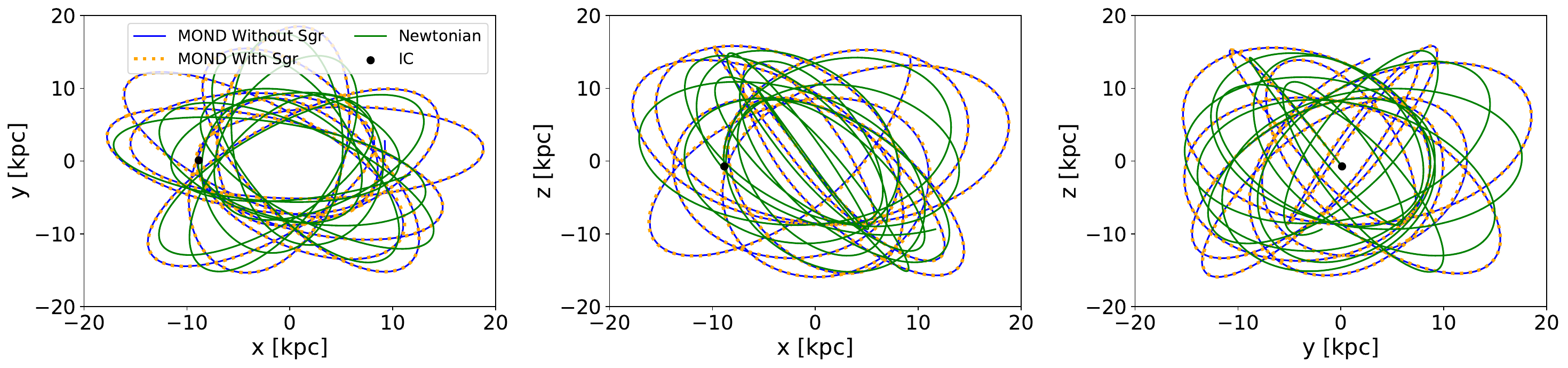}
    \caption{The orbit of one of the Helmi streams stars in the Newtonian potential (solid, green), the MOND potential (solid, blue) and the MOND potential including Sgr (dotted, orange) for the PW model. The dot shows the initial position of all three orbits.} 
    \label{fig:orbits}
\end{figure*}

For the LMC and Sgr, using a similar argument we find that their influence evolves over time while they orbit the MW, and that they enter regimes where their influence on the Helmi stream stars is not negligible \citep[see also ][]{Brada2000}. Therefore we choose to run additional tests where we include Sgr or the LMC while integrating the Helmi stream stars orbits. To this end we model both Sgr and the LMC using Hernquist spherical potentials with $M_{\mathrm{Sgr}}=1\times10^8M_{\odot}$, $a_{\rm H,\mathrm{Sgr}}=0.5$~\si{kpc} \citep{Vasiliev2020,McConnachie2012}, and $M_{\mathrm{LMC}}=3.2\times10^9M_{\odot}$, $a_{\rm H,\mathrm{LMC}}=2$~\si{kpc}  \citep{Besla2015}, where their masses correspond to estimates of their baryonic content. 

\subsection{Simplified Orbit Integrations}
As a first step to investigate the Helmi streams in MOND, we performed orbit integrations in the MOND framework,  {simplified by omitting the influence of the curl field $\bm{S}$} and compare these with orbits integrated in the original McM and PW Milky Way models that include dark matter halos and use Newtonian gravity.  {We note that we can only} safely omit the curl field $\bm{S}$ from our analysis and use $\bm{g}_{\mathrm{MOND}}=\nu(|\bm{g}_{\rm N}|/a_0)\bm{g}_{\rm N}$ if one can use the relation $|\bm{\nabla}\Phi_N|=\Phi_{\rm N}^2/GM_{\rm b}$, where $M_{\rm b}$ is the baryonic mass of the model \citep{Brada1995,Milgrom1986}. Close to the disk where $|z|<1.5$~kpc (2.5~kpc for the McM model)  {and within $R\lesssim30~\si{kpc}$} the above relation  {does not hold}. The LMC does not pass this region on its orbit, so we can use Eq.~(\ref{eq:gmond}) for integrating its orbit.  {Sgr has a pericentre of $\sim 20$~kpc and the stars in our Helmi Streams sample are currently located within $2.5\si{kpc}$ from the Sun. Still, for the majority of the time the orbits of Sgr and the Helmi streams are in a region where our simplification holds. Reassuringly, in Section \ref{sec:PoR}, where we explore the behaviour of the Helmi Streams using PoR, we will see that their orbits are barely affected by our simplification.}

The  {validity of this approach for the orbit integrations outside of the disk area} does not change if we add Sgr and the LMC to the MOND Milky Way potential. Their contribution is both small and locally spherically symmetric, so that $\bm{S}$ can still be neglected since $|\bm{\nabla}\Phi_{\rm N}|=\Phi_{\rm N}^2/MG$ holds in the region of the Helmi streams at each time in the presence of either Sgr or the LMC.

To compute the orbit of a star from the Helmi streams, we use its present-day position and velocity and integrate the equation of motion backward in time using the MOND force associated to the Milky Way potential under consideration. We use a Leapfrog integrator, with a timestep of $\Delta t=0.2$~\si{Myr}. For the Newtonian models we use the integrator implemented in \texttt{AGAMA} with the same timestep. 

To include either of the Milky Way satellites, namely Sgr or the LMC, we first determine their orbit in the MOND Milky Way model without the Helmi stream stars. Then we use the appropriate Hernquist potential representing the baryonic component of the satellite at the position of the satellite through time as an addition to the background potential we integrate our Helmi stream stars in. Figure \ref{fig:orbits} shows the orbit of one of the Helmi streams stars in the PW model. For comparison we also plot the orbit obtained including Sgr, as well as that using Newtonian gravity. This figure shows that the presence of Sgr does not lead to significant differences in the trajectory of the Helmi stream stars but that there are important differences between the MOND and Newtonian frameworks. 

\begin{figure*}
    \centering
    \includegraphics[width=\linewidth]{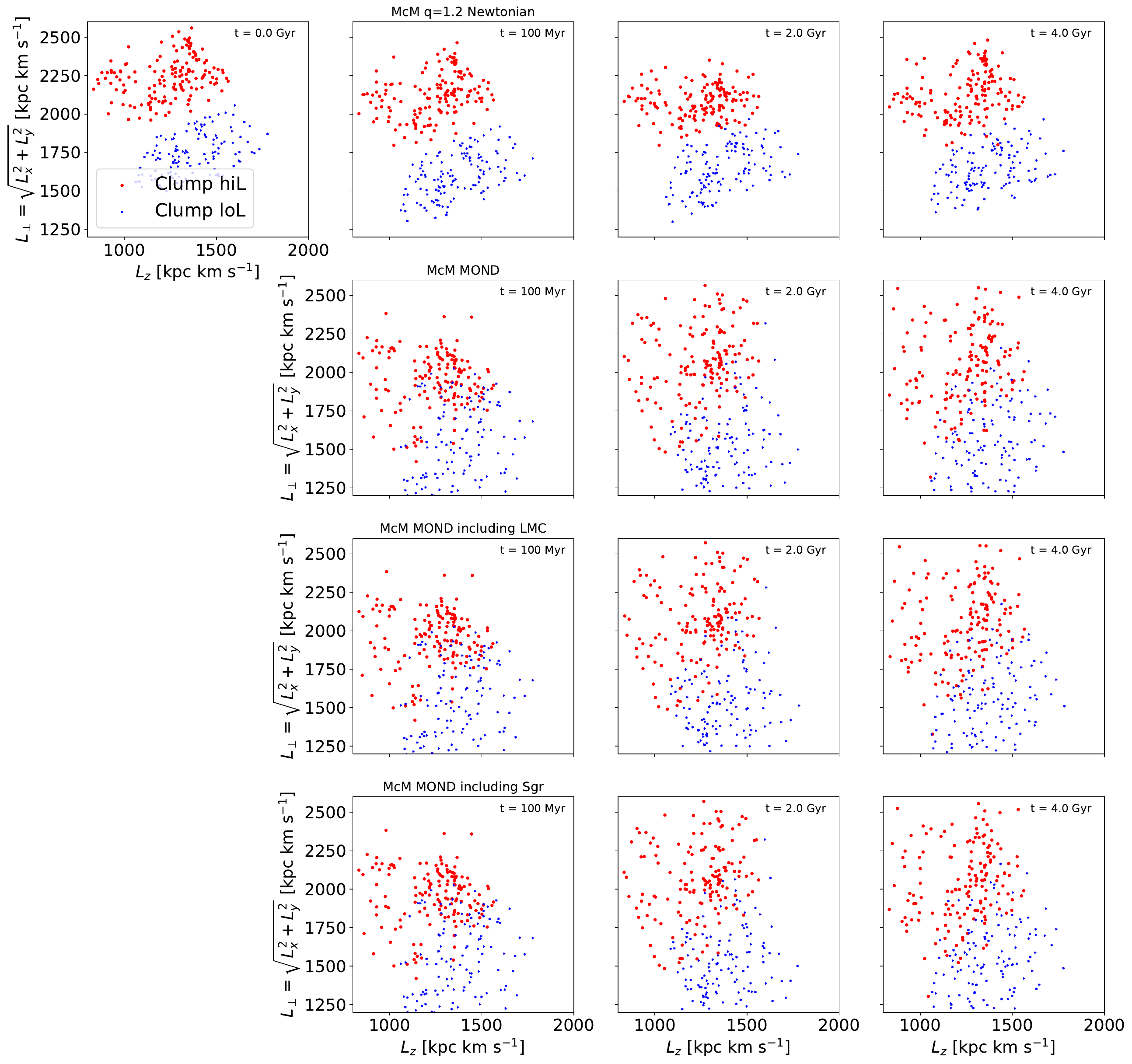}
    \caption{Evolution of the Helmi streams stars in $L_z-L_{\perp}$ over time in the  {McM Newtonian model of the Milky Way (with a modified dark matter halo with a flattening of $q=1.2$ as favored by D22, top row)}, the MOND version of that potential (second row) and the MOND versions including the effect of Sgr (third row) and of the LMC (bottom row). The first column shows the present day observed distribution, the second column the distribution after $\sim100$~\si{Myr} and the third and fourth column show the distribution after $\sim2$ and $\sim4$~\si{Gyr} of integration respectively.}
    \label{fig:results1}
\end{figure*}

\begin{figure*}
    \centering
    \includegraphics[width=\linewidth]{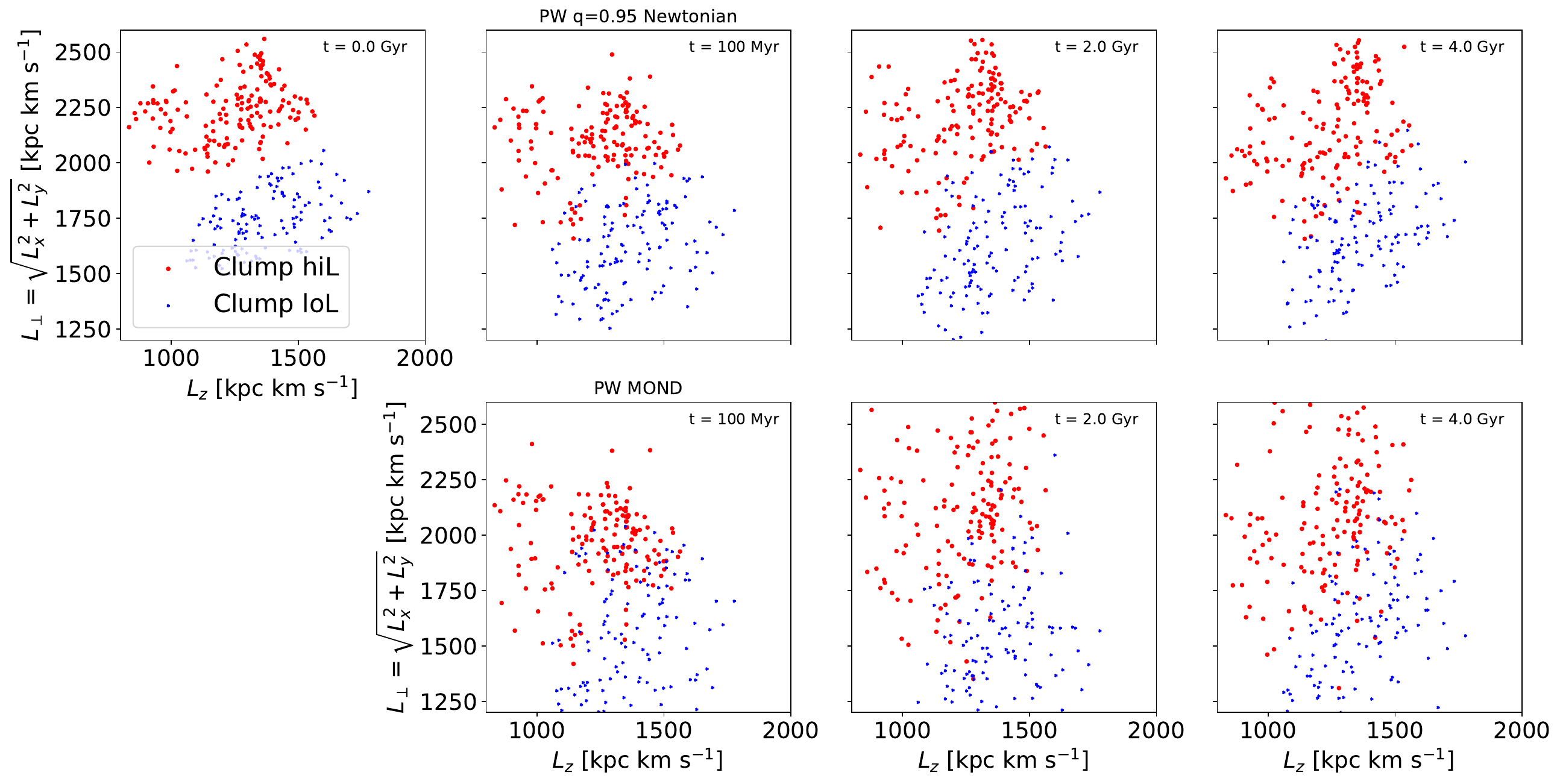}
    \caption{ {Same as Fig. \ref{fig:results1}, but for the PW model, where the top row shows the results for the Newtonian PW potential (which has a flattening of $q=0.95$).}}
    \label{fig:results2}
\end{figure*}

Figures \ref{fig:results1} and \ref{fig:results2} show the behaviour 
of the Helmi streams stars in the $L_z-L_{\perp}$ plane stemming from our  {simplified} orbit integrations, for the  {McM and PW} models respectively. In Fig. \ref{fig:results1}, the top row corresponds to the Newtonian model, the second row to the corresponding MOND model, and the bottom two rows include Sgr and the LMC. The columns show  {four} snapshots through the time of integration, at 0,  {0.1}, 2 and 4~Gyr.  {Because the results for adding Sgr and LMC to the PW model were similar to the results for the McM model, we do not show these in Fig. \ref{fig:results2} again.}

Clearly the gap between the two Helmi streams clumps does not persist in the MOND models, irrespective of the inclusion of Sgr or the LMC,  {which do not change the orbital behaviour visibly. The result is also valid} irrespective of the choice of baryonic potential for the Milky Way.\\

The two clumps mix significantly already within the first $\sim100$~\si{Myr} in all MOND models, with 
blue and red stars changing their $L_{\perp}$ enough that a distinction between the two groups is no longer possible. 
This extremely fast mixing behaviour is not seen for the Newtonian PW model that has an oblate dark matter halo with flattening of $q=0.95$. However, in this Newtonian model the gap does not persist either over long timescales as it disappears after $\sim200$~\si{Myr}, although the degree and rate by which the stars in the red and blue clumps are mixing is lower in comparison to the MOND models.  {In the Newtonian McM model with $q_{\rho}=1.2$ there is no mixing by construction (see D22 for details).}

%% file: 3PoR.tex
\section{Phantom of RAMSES}\label{sec:PoR}

 {We also calculate the orbits of the Helmi streams stars in the McM model using the Phantom of RAMSES (PoR) patch to the RAMSES N-body code to provide insight in a more sophisticated MOND framework \citep{Lughausen2015,Teyssier2010}.} 

 {We note that the PoR patch is based on the QUMOND formulation of MOND, whereas our orbit integrations and discussion of the $\bm{S}$ field were based on the AQUAL formulation of MOND. Hence, results between the two frameworks need not necessarily agree \citep{Milgrom2010,Famaey2012}. We employ the \texttt{staticpart} flag to represent the baryonic components of the McM model with a collection of particles that continually generate the MONDian gravitational potential the Helmi streams stars are integrated in. We generate the positions of these static particles using \texttt{AGAMA} \citep{Vasiliev2019} to represent the baryonic density profiles. We integrate the orbits of the Helmi streams stars for $2~\si{Gyr}$}{, using a minimum and maximum grid level of 7 and 18 respectively. This is sufficiently precise to model the orbit well. The boxsize of the simulation is $1024\si{kpc}$, ensuring that the potential can be approximated by a point source at the boundary of the box.}

\begin{figure*}
    \centering
    \includegraphics[width=\linewidth]{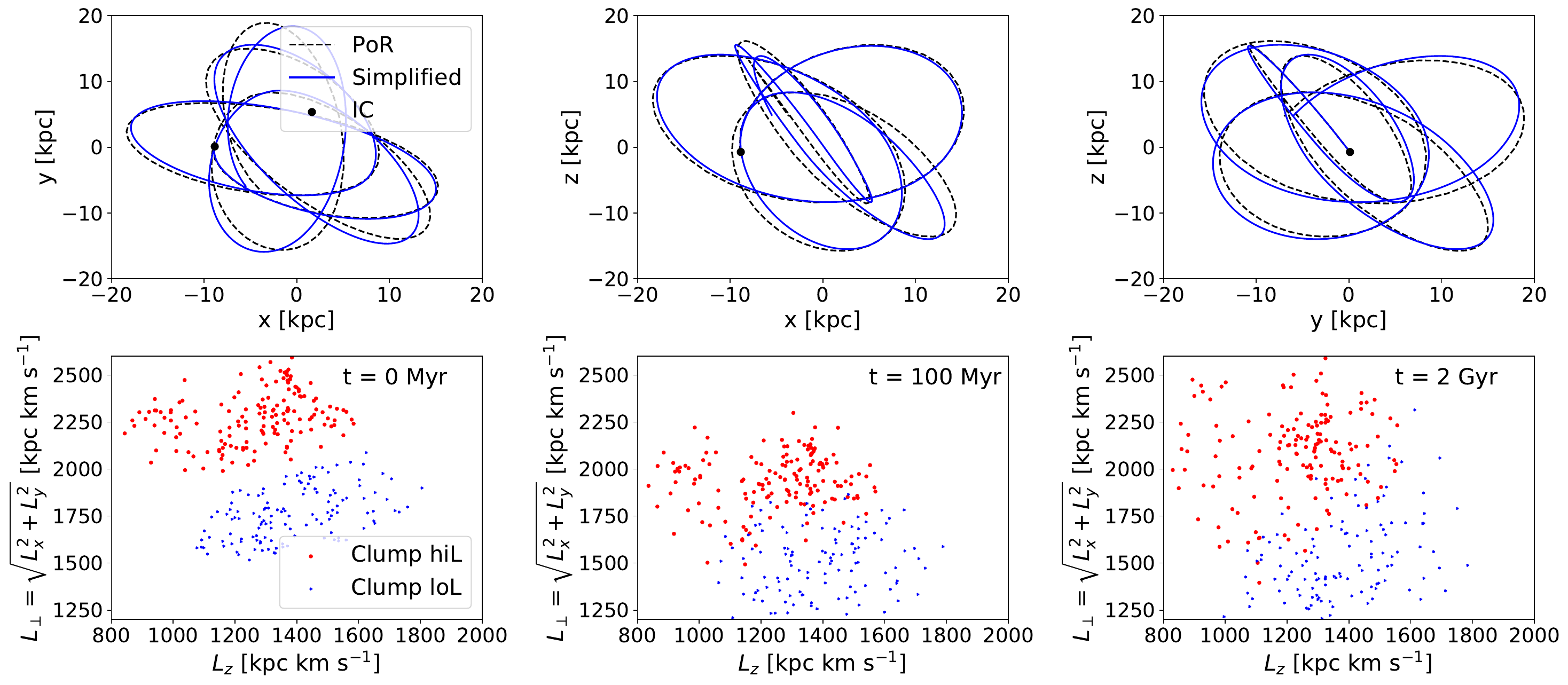}
    \caption{\textit{Top row:} The three projections of the orbit of one of the Helmi streams stars in the simplified MOND orbit integrations from Sec. \ref{sec:methods} (solid, blue) and in Phantom of RAMSES (dashed, black). The dot shows the initial position. \textit{Bottom row:} The $L_z-L_{\perp}$ plane for the Helmi streams stars in the PoR simulation at timestamps of (left to right) $0,0.1$ and $2~\si{Gyr}$.}
    \label{fig:results3}
\end{figure*}

 {The top row of Figure \ref{fig:results3} shows the orbit of one of the Helmi streams stars in the PoR simulation compared to its orbit in the simplified orbit integration from Section \ref{sec:methods}. Even though there is a difference that builds up over this time, the nature of the orbit is not altered. The bottom row shows the $L_z-L_{\perp}$ plane for the Helmi streams particles in the PoR simulation at timestamps of $t=0,0.1$ and $2~\si{Gyr}$. Again, we see that the gap between the two Helmi streams clumps dissolves within $100~\si{Myr}$, although the degree of mixing of the stars in the two clumps is less than in the simplified orbit integrations shown in the second row of Fig. \ref{fig:results1}. After $2~\si{Gyr}$ however, the two clumps show a similar degree of mixing in both the PoR and simplified orbit integrations.}

We may therefore conclude from Figs. \ref{fig:results1}, \ref{fig:results2}  {and \ref{fig:results3}} that the 
 {simplified} modelling  {as well as the sophisticated PoR integrations} of the Milky Way in MOND employed here fail to reproduce the
observations of the Helmi streams. Whether this is indicative of a shortcoming of our implementation
of MOND or that MOND theory, perhaps as a modification of inertia, should be altered, remains to be
seen.

%% file: 5Discussion_and_Conclusion.tex
\section{\label{sec:Discussion and Conclusion}Discussion and Conclusion}

We have shown by way of simplified orbit integrations in MOND models of the Milky Way using AQUAL  {as well as orbit integrations in QUMOND using the PoR patch for RAMSES} that the gap present between the two observed clumps in $L_z-L_{\perp}$ space for the Helmi streams does not persist through time. The stars in the clumps intermix very quickly, namely already in the first $\sim100$~\si{Myr}
of integration as can be seen in Figs. \ref{fig:results1}, \ref{fig:results2}  {and \ref{fig:results3}}, leading to the dissolution of the gap  {and the mixing of the clumps}. If we take our implementation of MOND at face value, this puts us in the uncomfortable position of having 
to argue that we are living at a truly special time in Galactic
history. 

A question that may naturally arise is whether the presence of two separate clumps could be reflecting some internal
substructure already existing in the progenitor system. And in that case, whether the gap between the clumps was initially larger such that it could have developed to its present observed form after a few Gyr of evolution in the Milky Way (although this may even be challenging for the MOND models). There is however a fundamental problem with this argument. By definition, substructure is localized in phase-space in a (progenitor) system. As the system is accreted, phase-space folds leading to the presence of multiple streams. Each of the streams stems from a (very) small region of the original system \citep[particularly if phase-mixing has occurred as is the case for the Helmi streams,][]{Helmi1999a}. Therefore, substructure in the progenitor system should be apparent in a localized portion of a stream.  Fig.~\ref{fig:data} shows that multiple kinematic streams (e.g. with positive and negative $v_z$) are associated to each of the two clumps in angular momentum space. This indicates that the clumps cannot be the result a local feature/substructure in the progenitor system, as stars in a given clump stem from a large region of phase-space of this system.  

 {Our results can
be compared to Figs. 8 and 9 of D22, who have shown that for the clumps to be distinguishable over long timescales, a prolate halo with a density flattening $q_{\rho} \sim 1.2$ would seem to be necessary, as we have exemplified in the top panels of Fig.~\ref{fig:results1}. This renders the effective potential to be close to spherical somewhere in the region probed by the stream stars, which results in better conservation of $L_{\perp}$, and hence of the clumps individuality. Such a prolate shape for the dark matter halo cannot be naturally produced in a MOND framework, since the gravitational field is
constrained to follow the shape of the baryonic distribution, which is very oblate (see Fig.~\ref{fig:densities}). The failure of our MONDian orbit integrations in reproducing the observed properties of the streams is thus not so surprising. }

 {We expect our conclusions to be robust 
since we have shown here that the dissolution of the gap between the clumps happens both in AQUAL with simplified orbit integrations either with or without the influence of the two biggest satellites of the Milky Way, as well as in the QUMOND framework studied through the PoR patch for RAMSES}. 
We have also run our integrations using a Kuzmin model for the Milky Way potential as in \cite{Read2005}, where $\bm{S}=0$ holds strictly, and the results were not different from those presented here.

 {Nonetheless it would be important to explore other possible formulations of MOND,
and specifically as a modification of inertia}, which unfortunately is not completely developed yet. Complications arising in this framework are, for instance, the non-locality in time and the fact that the definition of actions and conserved quantities like momentum could be different from the Newtonian definitions \citep{Milgrom2005,Milgrom2022}. 

The observations of the Helmi streams clumps
provide a particularly good test for MOND on Galactic scales, because the stars probe a
region governed mainly by the gravitational field of the Galaxy. Furthermore, their orbits are elongated in the direction perpendicular to the Galactic disk, a regime that has not been explored much in the context of MOND. As discussed in e.g. \cite{Zhu2023}, MOND and Newtonian gravity including a dark matter halo are able to fit equally well the Galaxy's rotation curve. There is thus a need for more studies like ours that probe a larger distance from the disk plane. An interesting example is the proposal to use high-velocity stars, as these objects probe regions where the two theories would predict rather different behaviour \citep{Chakrabarty2022}.
\\